\documentclass[aps,prc,twocolumn,showpacs,superscriptaddress,groupedaddress]{revtex4}  
\usepackage{graphicx}  
\usepackage{dcolumn}   
\usepackage{bm}        
\usepackage{amssymb}   
\usepackage{amsmath}

\begin{document}

\title{The full weak charge density distribution of $^{48}$Ca from parity violating electron scattering}
\author{Z. Lin}
\author{C. J. Horowitz}       
\affiliation{Center for the Exploration of Energy and Matter and Department of Physics, Indiana University, Bloomington, IN 47405, USA}                             
\date{\today}

\begin{abstract}
\noindent
{\bf Background:} The ground state neutron density of a medium mass nucleus contains fundamental nuclear structure information and is at present relatively poorly known.

\noindent
{\bf Purpose:} We explore if parity violating elastic electron scattering can provide a feasible and model independent way to determine not just the neutron radius but the full radial shape of the neutron density $\rho_n(r)$ and the weak charge density $\rho_W(r)$ of a nucleus.

\noindent
{\bf Methods:} We expand the weak charge density of $^{48}$Ca in a model independent Fourier Bessel series and calculate the statistical errors in the individual coefficients that might be obtainable in a model parity violating electron scattering experiment.

\noindent
{\bf Results:} We find that it is feasible to determine roughly six Fourier Bessel coefficients of the weak charge density of $^{48}$Ca within a reasonable amount of beam time.  However, it would likely be much harder to determine the full weak density of a significantly heavier nucleus such as $^{208}$Pb.

\noindent
{\bf Conclusions:} Parity violating elastic electron scattering can determine the full weak charge density of a medium mass nucleus in a model independent way.  This weak density contains fundamental information on the size, surface thickness, shell oscillations, and saturation density of the neutron distribution in a nucleus.  The measured $\rho_W(r)$, combined with the previously known charge density $\rho_{ch}(r)$, will literally provide a detailed textbook picture of where the neutrons and protons are located in an atomic nucleus.

\end{abstract}

\pacs{25.30.Bf, 27.40.+z, 21.10.Gv, 21.10.Ft}
\maketitle

\section{Introduction}

Where are the protons located in an atomic nucleus?  Historically, charge densities from elastic electron scattering have provided accurate and model independent information \cite{1}.  These densities are, quite literally, our  picture of the nucleus and have had an enormous impact.  They have helped reveal the size, surface thickness, shell structure, and saturation density of nuclei.    

Where are the neutrons located in an atomic nucleus?  Additional, very fundamental, nuclear structure information could be extracted if we also had accurate neutron densities.  For example, knowing both the proton and the neutron densities would provide constraints on the isovector channel of the nuclear effective interaction, which is essential for the structure of very neutron rich exotic nuclei.

However, compared to charge densities, our present knowledge of neutron densities is relatively poor and may be model dependent.  Often neutron densities are determined with strongly interacting probes \cite{2} such as antiprotons \cite{3,4}, proton elastic scattering \cite{5}, heavy ion collisions \cite{7}, Pion elastic scattering \cite{8}, and coherent pion photo production \cite{9}.  Here one typically measures the convolution of the neutron density with an effective strong interaction range for the probe.  Uncertainties in this range, from complexities of the strong interactions, can introduce significant systematic errors in the extracted neutron densities. 

It is also possible to measure neutron densities with electro-weak interactions, by using neutrino-nucleus coherent scattering \cite{10,11} or parity violating electron scattering \cite{20}.   This is because the weak charge of a neutron is much larger than that of a proton.  Compared to strongly interacting probes, parity violation provides a clean and model-independent way to determine the neutron density and likely has much smaller strong interaction uncertainties.  In the last decades, great theoretical \cite{12,13,14,15,16,17,18}  and experimental \cite{19,20} efforts have been made to improve parity violating electron scattering experiments.  At Jefferson laboratory, the neutron radius of $^{208}$Pb has been preliminarily measured by PREX \cite{20,22}, and will be measured with higher accuracy by the PREX-II  experiment \cite{PREXII}, while an approved experiment CREX aims to measure the neutron radius of $^{48}$Ca \cite{CREX}. 

In this paper, we propose to measure not only the neutron radius, but the full radial structure of the weak charge density distribution in $^{48}$Ca, by measuring the parity violating asymmetry at a number of different momentum transfers.  This will determine the coefficients of a Fourier Bessel expansion of the weak charge density that is model independent.  By measuring the weak density, the full structure of neutron density can be derived, since the weak form factor of the neutron is largely known and the weak charge of the proton is very small.   Our formalism to determine the cross-section for longitudinally polarized electrons scattered from $^{48}$Ca and the parity violating asymmetry $A_{pv}$ is presented in Section II.  In Section III we motivate measuring the full radial dependence of the weak charge density in $^{48}$Ca and discuss the large information that it contains.  In Section IV we illustrate our formalism with an example experiment and calculate the resulting statistical errors.  The resulting weak density can be directly compared to modern microscopic calculations of the ground state structure of $^{48}$Ca using Chiral effective field theory interactions \cite{cc48Ca}.  We conclude in Section V that it is feasible to measure the full weak density distribution of $^{48}$Ca.  However this may be much harder for a significantly heavier nucleus such as $^{208}$Pb because more Fourier Bessel coefficients likely will be needed.    

\section{Formalism}
\begin{figure}[tbf]
 \includegraphics[scale=0.33]{{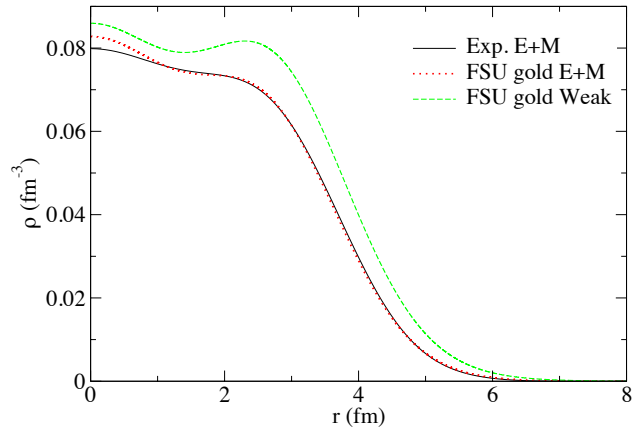}}
 \caption{\label{fig:2} (Color online) Ground state (electromagnetic) charge and weak charge densities of $^{48}$Ca versus radius $r$. The solid black line shows the Fourier Bessel experimental charge density from Ref. \cite{1} while the dotted red line shows the  charge density of the FSU-Gold relativistic mean field model.  Finally the green dashed line shows the weak charge density of the FSU-Gold model.}
 \end{figure}

The parity violating asymmetry for longitudinally polarized electrons scattering from a spin zero nucleus, $A_{pv}$, is the key observable which is very sensitive to the weak charge distribution.  The close relationship between $A_{pv}$ and the weak charge density $\rho_W(r)$ can be readily seen in the Born approximation, 
\begin{equation} 
A_{pv}\equiv\frac{d\sigma/d\Omega_R+ d\sigma/d\Omega_L}{d\sigma/d\Omega_R - d\sigma/d\Omega_L}\approx-\frac{G_Fq^2}{4\pi\alpha\sqrt{2}}\frac{Q_WF_W(q^2)}{ZF_{ch}(q^2)}.
\label{eq:1}
\end{equation}
Here $d\sigma/d\Omega_R$ ($d\sigma/d\Omega_L$) is the cross section for positive (negative) helicity electrons, $G_F$ is the Fermi constant, $q$ the momentum transfer, $\alpha$ the fine structure constant, and $F_W(q^2)$ and $F_{ch}(q^2)$ are the weak and charge form factors respectively, 
\begin{equation}
F_W(q^2)=\frac{1}{Q_W}\int d^3r j_0(qr) \rho_{W}(r)
\label{eq:2}
\end{equation}
\begin{equation}
F_{ch}(q^2)=\frac{1}{Z}\int d^3r j_0(qr)\rho_{ch}(r).
\label{eq:3}
\end{equation}
These are normalized $F_W(0)=F_{ch}(0)=1$.  The charge density is $\rho_{ch}(r)$ and $Z=\int d^3r \rho_{ch}(r)$ is the total charge.  Finally, the weak charge density $\rho_W(r)$, see Fig. \ref{fig:2}, and the total weak charge $Q_W=\int d^3r \rho_W(r)$ are discussed below.

\begin{figure}[tbf]
 \includegraphics[scale=0.33]{{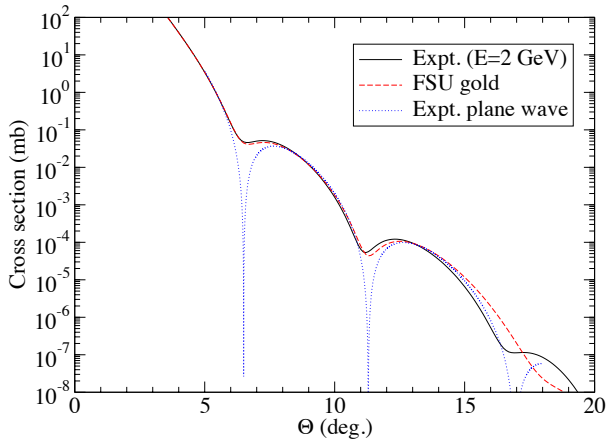}}
 \caption{\label{fig:1} (Color online) Differential cross section for 2 GeV electrons elastically scattered from $^{48}$Ca versus scattering angle. The experimental charge density is used for the solid black line including Coulomb distortions and the dotted blue line in a plane wave impulse approximation.  Finally the dashed red line uses the model FSU Gold relativistic mean field theory charge density including Coulomb distortions. }
 \end{figure}

The elastic cross-section in the plane wave Born approximation is,
\begin{equation}
\frac{d\sigma}{d\Omega}=\frac{\alpha^2\cos^2 (\frac{\theta}{2})}{4E^2\sin^4 (\frac{\theta}{2})}\bigl|F_{ch}(q^2)\bigr|^2,
\label{eq:4}
\end{equation} 
with $\theta$ the scattering angle.  However, for a heavy nucleus, Coulomb-distortion effects must be included \cite{12}.  In Fig. \ref{fig:1} we compare the plane-wave cross-section, Eq. \ref{eq:4}, to the cross section including Coulomb-distortion effects, see for example \cite{17}. Coulomb distortions are seen to fill in the diffraction minima.  However away from these minima distortion effects on the cross section are relatively small.  The cross section calculated with the charge density from a relativistic mean filed model using the FSU-gold interaction \cite{FSUGold}, see Fig. \ref{fig:2}, agrees well with the experimental charge density except at the largest angles.
 
We now expand the weak density of $^{48}$Ca in a Fourier Bessel series.  We truncate this expansion after $n_{max}$ terms and  assume the weak density $\rho_W(r)$ is zero for r$>R_{max}$.  This expansion will be model independent if truncation errors are small.  
\begin{equation}
\rho_{W}(r)=\sum_{i=1}^{n_{max}}a_ij_0(q_ir)
\label{eq:5}
\end{equation} 
Here $q_i$=${i\pi}/{R_{max}}$ and $j_{0}(x)=\sin(x)/x$.

To minimize measurement time we would like $n_{max}$ and $R_{max}$ to be as small as possible while still accurately representing the full weak density.  In this paper we consider
\begin{equation}
R_{max}=7\ {\rm fm},
\end{equation}
since the weak charge density determined from many density functionals is small for $r>$ 7 fm.  In addition we use,
\begin{equation}
n_{max}=6,
\end{equation} 
because the expansion coefficients $a_i$ determined for many density functional calculations of $^{48}$Ca are very small for $i>6$.  We determine truncation errors using a model weak charge density based on the FSU Gold relativistic mean field interaction \cite{FSUGold}, see below.  This model density has $1.8\times 10^{-3}$ of the weak charge at $r>R_{max}=7$ fm, and the expansion coefficients $|a_i|$ for $i>n_{max}=6$ are all $<7\times 10^{-4}$ fm$^{-3}$.  This is an order of magnitude or more smaller than the smallest $|a_i|$ for $i\le n_{max}$.

We now consider determining the six coefficients $a_i$ for $i=1$ to 6.  In plane wave Born approximation, a given $a_j$ can be determined from a measurement of  $A_{pv}(q_j)$ at  momentum transfer $q_j=j\pi/R_{max}$.    In principle only five measurements are needed to determine the six $a_i$ because the weak density is normalized to the total weak charge $\int d^3r \rho_W(r)=Q_W$.  To be very conservative we use this normalization condition to determine $a_1$.   If instead we used the normalization to determine $a_6$, considerably less beam time might be needed for a given statistical accuracy.  However, the resulting density might then be more sensitive to truncation errors.  

Note that in plane wave Born approximation $A_{pv}(q_j)$ is only sensitive to $a_j$ because of the orthogonality of the Fourier Bessel series.  When Coulomb distortions are included $A_{pv}(q_j)$ is still primarily sensitive to $a_j$ and only depends very slightly on the other coefficients. This will be shown in Figs. \ref{fig:4} and \ref{fig:5} in Section IV.  

We consider a reference weak charge density by calculating $\rho_W(r)$ for a realistic model and then expanding the model density in the Fourier Bessel series.  For the model the expansion coefficients are given by, 
\begin{equation}
a_i=\frac{\int_0^{R_{max}}\rho_W(r)j_0(q_ir)r^2\,dr}{\int_0^{R_{max}}j_0^2(q_ir)r^2\,dr}\, .
\label{eq:6}
\end{equation} 
After having determined the weak density distribution, we can constrain the neutron density $\rho_n(r)$ in $^{48}$Ca since the neutron density is closely related to the weak density.  If one neglects spin-orbit currents that are discussed in Ref. \cite{spin_orbit}, and other meson exchange currents \cite{big_paper} one can write,
\begin{equation}
\rho_{W}(r)=\int d^3r' \{4G_n^Z(\lvert{r-r'}\rvert)\rho_n(r)+4G_p^Z(\lvert{r-r'}\rvert)\rho_p(r)\}\, .
\label{eq:7}
\end{equation}
Here $\rho_p(r)$ is the proton density and $G_n^Z(r)$ and $G_p^Z(r)$ are the Fourier transforms of the neutron and proton single nucleon weak form factors,  
\begin{equation}
4G_n^Z(r)=Q_nG_E^p(r)+Q_pG_E^n(r)-G_E^s(r),
\end{equation}
\begin{equation}
4G_p^Z(r)=Q_pG_E^p(r)+Q_nG_E^n(r)-G_E^s(r),
\end{equation}
where $G_E^p(r)$ and $G_E^n(r)$ are Fourier transforms of the proton and neutron electric form factors.  They are normalized
\begin{equation}
\int d^3r G_E^p(r)=1, \ \ \ \ \ \ \ \ \int d^3r G_E^n(r)=0.
\end{equation}
Finally $G_E^s(r)$ is the Fourier transform of strange quark contributions to the nucleon electric form factor \cite{strange1,strange2,strange3,strange4} and is normalized $\int d^3r G_E^s(r)=0$.
The weak form factors are normalized,
\begin{equation}
\int d^3r\, 4G_n^Z(r)= Q_n, \ \ \ \ \ \ \ \int d^3r\, 4G_p^Z(r)=Q_p.
\end{equation}
The weak charge of the neutron $Q_n$ is -1 at tree level, while the weak charge of the proton $Q_p$ is $1-4\sin^2\theta_W$ at tree level.  Including radiative corrections \cite{rad1,rad2} one has,
\begin{equation}
Q_n=-0.9878, \ \ \ \ \ \ \ \ \ Q_p=0.0721\, .
\end{equation}
Finally, the total weak charge of $^{48}$Ca is,
\begin{equation}
Q_W=\int d^3r \rho_W(r)=NQ_n+ZQ_p=-26.216\,.
\end{equation}
Further radiative corrections, for example from $\gamma-Z$ box diagrams \cite{gzbox1,gzbox2}, are not expected to be important compared to this large value of $Q_W$.

We emphasize that parity violating experiments can determine the weak density $\rho_W(r)$ in a model independent fashion.  This can be compared to theoretical predictions for $\rho_W(r)$ that are obtained by folding theoretical nucleon densities $\rho_n(r)$ and $\rho_p(r)$ with single nucleon weak form factors and possibly including meson exchange current contributions.

\section{Motivation}
In this section we discuss the information content in the weak charge density and some of the physics that would be constrained by measuring $\rho_W(r)$ with parity violating electron scattering.  First the weak radius $R_w=[\int d^3r r^2 \rho_W(r)/Q_W]^{1/2}$ is closely related to the neutron radius, see for example \cite{22}.  This has been extensively discussed.  

The surface thickness of $\rho_W(r)$ can differ from the known surface thickness of $\rho_{ch}(r)$ and is expected to be sensitive to poorly constrained isovector gradient terms in energy functionals.  One way to constrain these gradient terms is to perform microscopic calculations of pure neutron drops in artificial external potentials, using two and three neutron forces. Then one can fit the resulting energies and neutron density distributions with an energy functional by adjusting the isovector gradient terms.  It may be possible to test these theoretically constrained isovector gradient terms by measuring the surface behavior of $\rho_W(r)$.

Next, the interior value of $\rho_W(r)$ for small $r$ is closely related to the interior neutron density.  This, when combined with the known charge density, will finally provide a direct measurement of the interior baryon density of a medium mass nucleus.  Previously, this has only been extracted in model dependent ways by fitting a density functional to the charge density and then using the functional to calculate the baryon density.  This is sensitive to the form of the symmetry energy contained in the functional.  The interior baryon density is thought to saturate (stay approximately constant) with increasing mass number $A$.  This saturation density is closely related to the saturation density of infinite nuclear matter and insures that nuclear sizes scale approximately with $A^{1/3}$.

The saturation density of infinite nuclear matter $\rho_0\approx 0.16$ fm$^{-3}$ is a very fundamental nuclear property that has proved difficult to calculate.  It is very sensitive to three nucleon forces and calculations with only two body forces can saturate at more than twice $\rho_0$.  Indeed microscopic calculations that use phenomenological two nucleon forces fit to nucleon-nucleon scattering data and phenomenological three nucleon forces fit to properties of light nuclei may not be able to make sharp predictions for $\rho_0$ because of unconstrained short range behavior of the three nucleon forces \cite{23}.  As a result these calculations often do not predict $\rho_0$ but instead fit $\rho_0$ by adjusting three nucleon force parameters. 

Alternatively, chiral effective field theory provides a framework to expand two, three, and four or more nucleon forces in powers of momentum over a chiral scale.  There are now a growing number of calculations of nuclear matter, see for example \cite{24, 25}.  However, at this point it is unclear how well the chiral expansion convergences for symmetric matter at nuclear densities and above.  There could still be significant uncertainties from higher order terms in the chiral expansion and from the dependence of the calculation on the assumed cutoff parameter and on the form of regulators used.  Furthermore, there are uncertainties in the energy of nuclear matter that arise from uncertainties in short range parameters that are fit to other data.  Finally, some higher order terms in the chiral expansion may be somewhat larger than expected because of large contributions involving Delta baryons.

In addition to calculations for infinite nuclear matter, advances in computational techniques have now allowed improved microscopic calculations directly in finite nuclei.  These calculations can directly test nuclear saturation by seeing how predictions compare to data as a function of mass number $A$.   One can test both isoscalar and isovector parts of these calculations by comparison to both interior charge and weak charge densities.  The interior weak density may be sensitive to three neutron forces and reproducing it may allow better predictions for very neutron rich medium mass nuclei where three neutron forces may also play a very important role.

Finally we expect shell oscillations in $\rho_W(r)$.  Shell oscillations have been observed in $\rho_{ch}(r)$ for a variety of nuclei.  For example there is a small increase in $\rho_{ch}(r)$ for $^{208}$Pb as $r\rightarrow 0$ due to the filling of the 3S proton state.  However observed shell oscillations in $\rho_{ch}$ are often much smaller than those predicted in many density functional calculations.  Indeed almost all density functional calculations over predict the small $r$ bump in $\rho_{ch}(r)$ for $^{208}$Pb, see for example \cite{26}.  It may be very useful to finally have direct information on shell oscillations for neutrons in addition to protons.  This could suggest changes in the form of the density functionals that are used that would correct the shell oscillations.                 

We conclude this section.  A model independent determination of $\rho_W(r)$ and features of the neutron density including surface thickens, central value, and shell oscillations address a number of important current problems in nuclear physics.  Together with $\rho_{ch}(r)$ they will literally provide a detailed picture of where the neutrons and protons are in an atomic nucleus.

\section{Sample Experiment}

In this section we evaluate the statistical error for the measurement of the weak charge density of $^{48}$Ca in a sample experiment.   As an example we consider measuring $A_{pv}$ at five $q^2$ points during a single run in Hall A at Jefferson Laboratory.    The total measurement time for all five of the points is assumed to be 60 days. The experimental parameters including beam current $I$, beam polarization $P$, detector solid angle $\Delta\Omega$, number of arms $N$, and the radiation loss factor $\zeta$ are assumed to be similar to the CREX experiment \cite{CREX} and are listed in Table \ref{tab:table1}, see also Ref. \cite{28}.   This example provides a conservative baseline for the final statistical error.   Measuring some (or all) of the points at other laboratories such as Mainz, or combining data from other experiments could significantly reduce the statistical error.   For example if the CREX experiment is run first and provides a very accurate low $q^2$ point,  then that information could be used to reduce the number of future measurements needed to determine the full weak density.   We emphasize that in this section we only present statistical errors.  Of course any real experiment will also have systematic errors that we discuss briefly at the end of this section.

\begin{table}
\begin{ruledtabular}
\begin{tabular}{cc}
     Parameter Value\\
\hline
$I$ & 150$\mu$A\\
$P$ & 0.9 \\
$\rho_{\rm tar}$ & $2.4 \times 10^{22}$ cm$^{-2}$\\
$\Delta\Omega$ & 0.0037 Sr \\
N & 2\\
$\zeta$ & 0.34\\
\end{tabular}
\end{ruledtabular}
\caption{\label{tab:table1}Assumed experimental parameters including beam current
$I$, beam polarization $P$, target thickness $\rho_{\rm tar}$, detector solid angle $\Delta\Omega$, number of
arms N, and radiation loss factor $\zeta$. }
\end{table}

\begin{figure}[ftb]
  \includegraphics[scale=0.33]{{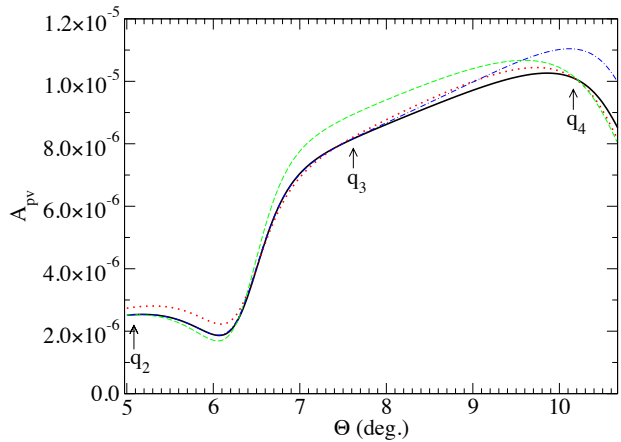}}
  \caption{\label{fig:4} (Color online) Parity violating asymmetry $A_{pv}$ for 2 GeV electrons elastically scattered from $^{48}$Ca versus scattering angle. The solid black line shows the asymmetry curve based on the charge density from Ref. \cite{1} and the FSU-Gold weak density. The red dotted, green dashed, and blue dot-dashed curves show $A_{pv}$ when $a_2$, $a_3$ and $a_4$ have been varied one at a time by ten percent respectively.  The arrows show the momentum transfers $q_i$ for $i=2,3$ and 4. }
  \end{figure}

\begin{figure}[ftb]
  \includegraphics[scale=0.33]{{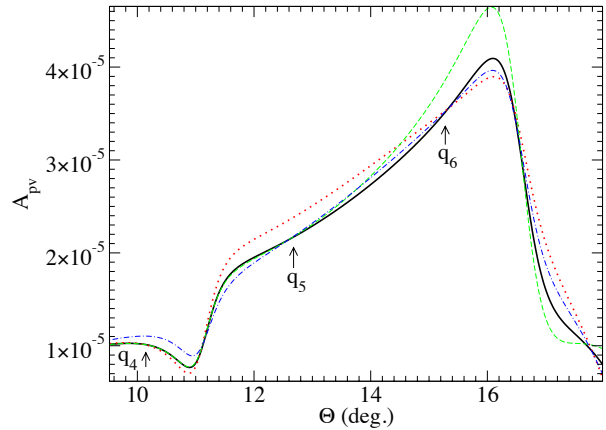}}
  \caption{\label{fig:5} (Color online) As per Fig. \ref{fig:4} but for larger scattering angles.  The arrows show $q_i$ for $i=4,5$ and 6.}
  \end{figure}

We calculate the statistical error in the determination of the Fourier Bessel coefficients $a_i$ of the weak density.  The total number of electrons detected $N_{tot}$ in a measurement time $T_i$ is 
\begin{equation}
N_{tot}=IT_i\rho_{tar}\frac{d\sigma}{d\Omega}\zeta\Delta\Omega N\, .
\label{eq:8}
\end{equation}
The statistical error in the determination of $a_i$ is $\Delta a_i$,
\begin{equation}
\frac{\Delta a_i}{a_i}=\Bigl(N_{tot}{A_{pv}(q_i)}^2{P}^2{\epsilon_i}^2\Bigr)^{-\frac{1}{2}}
\label{eq:9}
\end{equation}
Here $\epsilon_i$ is the sensitivity of $A_{pv}$ to a change in $a_i$ and is defined as,
\begin{equation}
\epsilon_i=\frac{\partial\ln A_{pv}(q_i)}{\partial\ln a_i}=\frac{a_i}{A_{pv}}\frac{\partial A_{pv}}{\partial a_i}\, .
\label{eq:10}
\end{equation}
In plane wave Born approximation $\epsilon_i=1$ and including Coulomb distortions one still has $\epsilon_i\approx 1$, see Figs. \ref{fig:4} and \ref{fig:5}.

Our reference weak charge density $\rho_W(r)$ for $^{48}$Ca is the Fourier Bessel expansion of a relativistic mean field theory model using the FSU-Gold interaction, see Fig. \ref{fig:2}. The Fourier Bessel coefficients determined from Eq. \ref{eq:6} are listed in Table II.  This model yields a charge density  for $^{48}$Ca that agrees well with the experimental charge density from Ref. \cite{1} except in the central region as shown in Fig. \ref{fig:2}. 
\begin{figure}[tbf]
 \includegraphics[scale=0.33]{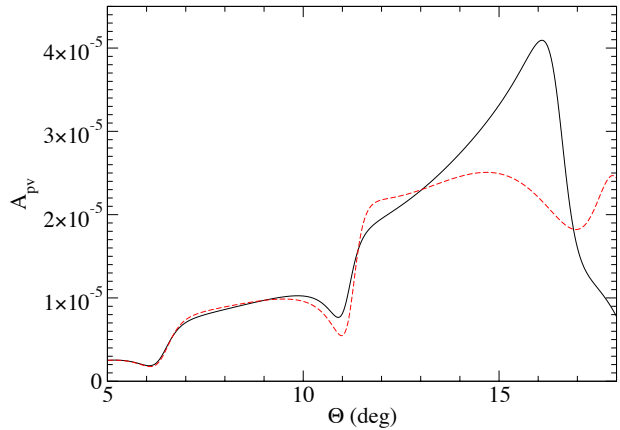}
 \caption{\label{fig:3} (Color online) Parity violating asymmetry $A_{pv}$ for 2 GeV electrons elastically scattered from $^{48}Ca$ versus scattering angle. The dashed red line is based on the weak and charge densities from the FSU-Gold relativistic mean field model. The solid black line uses the experimental charge density from Ref. \cite{1} and the FSU-Gold weak density.}
 \end{figure}

It is crucial to use a very accurate charge density in the determination of the weak charge density.  In Fig. \ref{fig:3} we show $A_{pv}$  calculated with the same reference weak density but with the FSU-Gold model or the experimental charge density.  There is a significant difference at large momentum transfers.  Note both curves include Coulomb distortions.  In the following we will always include Coulomb distortions using the code ELASTC \cite{12} and use the full experimental charge density.

 We now consider five measurements of $A_{pv}(q_i)$ at momentum transfers $q_i$ for $i=2$ to 6.  In general one may be able to improve statistics, for a given $q_i$, by going to a more forward angle and higher beam energy.  We somewhat arbitrarily restrict the scattering angle to be at least five degrees, since this is the scattering angle for the septum magnet of the PREX experiment.   We also limit the beam energy to no more than 4 GeV as a possible restriction from the HRS spectrometers in Hall A.  Thus we consider five measurements with the kinematics in Table \ref{tab:table2}.  

The statistical error in the total weak charge density $\rho_W(r)$ is the quadratic combination of the errors for the six Fourier Bessel terms.  Note that the statistical errors $\Delta a_i$ for different $i$ are independent.
\begin{eqnarray}
\Delta\rho_{weak}(r)=\Bigl\{\sum_{i=1}^{n_{max}}\Bigl[\Delta a_ij_0(q_ir)\Bigr]^2\Bigr\}^{\frac{1}{2}}
\label{eq:11}
\end{eqnarray}
The individual errors $\Delta a_i$ depend on the time $T_i$ spent measuring $A_{pv}(q_i)$ at momentum transfer $q_i$.  As a simple example we optimize the individual $T_i$, subject to a constraint on the total measurement time
\begin{equation}
\sum_{i=2}^6T_i=60\ {\rm days},
\end{equation}
in order to minimize the statistical error in $\rho_W(0)$ at $r=0$.   Note that the error in $a_1$ is calculated by using the normalization condition $\int \rho_{W}(r)d^3r$=$Q_{W}$.   The individual $T_i$ and the fractional errors in each of the $\Delta a_i$ are listed in Table \ref{tab:table2}.
\begin{table}
\begin{ruledtabular}
\begin{tabular}{ccccccccc}
 $q_i$ &  E   & $\frac{d\sigma}{d\Omega}$ & $A_{pv}$& T  &  $a_i$ & $\Delta a_i/a_i$ \\
  fm$^{-1}$  & GeV  & mb & ppm & days & fm$^{-3}$& \% \\
\hline
 0.45  &     &  &  &  & 0.0752 & $1.1$  \\
0.90  & 2.06 & 2.44 & 2.54& 5& 0.0468 &  $5.9$\\
1.35  & 3.09&  $1.07 \times 10^{-1}$& 8.31 & 7& -0.0438  & $7.6$\\
1.80  & 4 &  $2.9 \times 10^{-3}$ & 9.92 & 10& -0.0147& $27$\\
2.24 & 4 & $4.05\times 10^{-4}$ & 22.5& 15&0.0161 & $29$\\
2.69 & 4 & $9.7\times 10^{-6}$& 36.5& 23& 0.0066 &   $90$\\
\end{tabular}
\end{ruledtabular}
\caption{\label{tab:table2} The momentum transfer $q_i$, beam energy $E$, cross section, parity violating asymmetry $A_{pv}$, measurement time $T$,  Fourier Bessel expansion coefficient  $a_i$ of the weak charge density as determined for the FSU-gold relativistic mean field model, and fractional statistical error $\Delta a_i/a_i$.  Note that the error in $a_1$ is determined by normalizing the weak charge density to the total weak charge $Q_W$.}
\end{table}

Table \ref{tab:table2} shows that most of the time is spent measuring the highest momentum transfer points.  This is because the cross section falls so rapidly with increasing $q$.  One alternative, to this model independent approach, would be to constrain the higher $i$ coefficients $a_i$ from theory and only measure $A_{pv}$ for smaller momentum transfers.  This could significantly reduce the run time and the statistical error.


 
\begin{figure} 
   \includegraphics[scale=0.33]{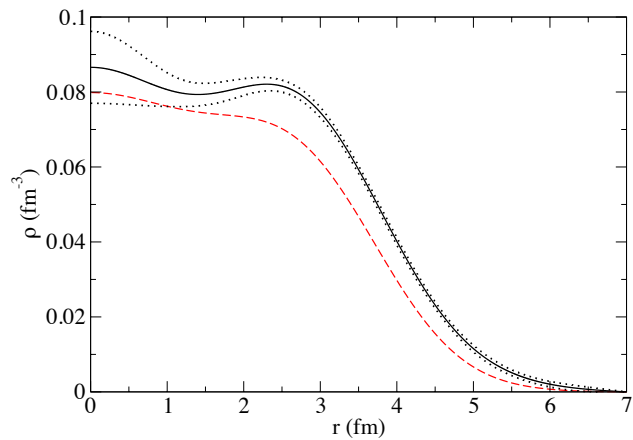}
   \caption{\label{fig:6} (Color Online) Weak charge density $\rho_W(r)$ of $^{48}$Ca versus radius $r$. The black solid line is the reference  FSU-Gold weak density and the black dotted lines show the statistical error band that could be obtained by measuring  $A_{pv}$ at five specific momentum transfers with a total running time of 60 days, see text.  The dashed red line shows the experimental (electromagnetic) charge density \cite{1}.}
   \end{figure}

Figure \ref{fig:6} shows the statistical error in the weak density $\rho_W(r)$ as a function of radius $r$. The error in $\rho_{W}$ is largest for small $r$ and gradually decreases as $r$ increases.  Thus it is most difficult to determine $\rho_W(r)$ near the origin.  There may be several ways to decrease the error band in Fig. \ref{fig:6}.  One could measure with higher beam currents and or with larger acceptance spectrometers.  Alternatively, one could measure for a larger time either as one extended experiment or by combining experiments that each focus on only some of the $q_i$ points.

We now discuss systematic errors.  Because it is so difficult to get good statistics for large $q$, the higher $i$ coefficients $a_i$ may only be determined with somewhat large statistical errors $\Delta a_i/a_i$.  As a result many systematic errors such as determining the absolute beam polarization or from helicity correlated beam properties may be less important.  Instead backgrounds, from for example electrons that scatter from collimators used to define the acceptance, could be important because the elastic cross section is small (at higher $q$).  

We have focused on $^{48}$Ca.  Determining the full $\rho_W(r)$ for a significantly heavier nucleus such as $^{208}$Pb may be dramatically harder.  This is because more Fourier Bessel coefficients will likely be needed and because the cross section drops extremly rapidly with increasing $q$.  Thus it may be very hard to measure $A_{pv}$, at high enough $q$, in order to directly determine the weak density in the center of $^{208}$Pb.

\section{Conclusions}

The ground state neutron density of a medium mass nucleus contains fundamental nuclear structure information and it is at present relatively poorly known.  In this paper we explored if parity violating elastic electron scattering can determine not just the neutron radius, but the entire radial form of the neutron density $\rho_n(r)$ or weak charge density $\rho_W(r)$ in a model independent way.  We expanded the weak charge density $\rho_W(r)$ in a model independent Fourier Bessel series.  For the medium mass neutron rich nucleus $^{48}$Ca, we find that a practical parity violating experiment could determine about six Fourier Bessel coefficients $a_i$ and thus deduce the full radial structure of both $\rho_W(r)$ and the neutron density $\rho_n(r)$.  The resulting $\rho_W(r)$ will contain fundamental information on the size, surface thickness, shell oscillations, and saturation density of the neutron distribution.

Future work could optimize our model experiment to further reduce the statistical errors by for example using large acceptance detectors and combining information from multiple experiments and or laboratories.   Future theoretical work exploring the range of weak charge densities to be expected with reasonable models and microscopic calculations would also be very useful.  The measured $\rho_W(r)$, combined with the previously known charge density $\rho_{ch}(r)$, will literally provide a detailed textbook picture of where the neutrons and protons are located in an atomic nucleus.

\section*{Acknowledgements}

We thank Bob Michaels for helpful comments and Shufang Ban for initial contributions to this work.  We thank the Mainz Institute for Theoretical Physics for their hospitality.  This research was supported in part by DOE grants DE-FG02-87ER40365 (Indiana University) and DE-SC0008808 (NUCLEI SciDAC Collaboration).


\end{document}